\documentclass[a4paper]{article}
\usepackage{dx}
\usepackage[T1]{fontenc}
\usepackage{ae,aecompl}
\usepackage{amsfonts}
\usepackage{amsmath}
\usepackage{amssymb}
\usepackage{amsthm}
\usepackage{cite}
\usepackage{float}
\usepackage{hyperref}       % hyperlinks
\usepackage{url}            % simple URL typesetting
\usepackage{booktabs}       % professional-quality tables
\usepackage{nicefrac}       % compact symbols for 1/2, etc.
\usepackage{graphicx}
\usepackage{doi}
\usepackage{dblfloatfix} 
\usepackage{times}
 \usepackage[frozencache=true,cachedir=minted-cache]{minted} 
\usepackage{balance}
\usepackage{flushend}
\theoremstyle{definition}
\newtheorem{definition}{Definition}[section]

\usepackage{fontawesome}

\title{On using distributed representations of source code for the \\ detection of C security vulnerabilities}
\author%
{%
David Coimbra\textsuperscript{\faCheck}, 
Sofia Reis\textsuperscript{\faCheck},
Rui Abreu\textsuperscript{\faCheckCircle},
Corina P\u{a}s\u{a}reanu\textsuperscript{\faCheckCircleO},
Hakan Erdogmus\textsuperscript{\faCheckCircleO} \\
\textsuperscript{\faCheck}INESC-ID \& IST/U.Lisbon, Portugal \\
\textsuperscript{\faCheckCircle}INESC-ID \& FEUP, Portugal \\
\textsuperscript{\faCheckCircleO}Carnegie Mellon University, USA \\
}

\begin{document}

\maketitle

\begin{abstract}
This paper presents an evaluation of the code representation model Code2vec when trained on the task of detecting security vulnerabilities in C source code. We leverage the open-source library astminer to extract path-contexts from the abstract syntax trees of a corpus of labeled C functions. Code2vec is trained on the resulting path-contexts with the task of classifying a function as vulnerable or non-vulnerable. Using the CodeXGLUE benchmark, we show that the accuracy of Code2vec for this task is comparable to simple transformer-based methods such as pre-trained RoBERTa, and outperforms more naive NLP-based methods. We achieved an accuracy of $61.43\%$ while maintaining low computational requirements relative to larger models. 
\end{abstract}

\section{Introduction}
\label{sec:intro}
Security vulnerabilities are a major concern in software development, as even the simplest mistakes can be turned into attack vectors by a malicious party. In continuous integration (CI), it is common to introduce static analysers into the build pipeline to verify code against known patterns~\cite{5504795, Kuusela2017, 7962383}. For example, Brakeman\footnote{\url{https://brakemanscanner.org/}} and SonarQube\footnote{\url{https://docs.sonarqube.org/latest/analysis/github-integration/}} are static analysers capable of detecting software vulnerabilities in source code that can be used for this purpose. 

Static analyzers can reduce security concerns in the software development lifecycle when introduced in the implementation phase, to assist developers to produce safer software. These analyzers are also useful tools for code review. Nowadays, one common practice to provide feedback is using GitHub pull requests. Warnings are usually added automatically to pull requests as comments, which can be reviewed manually before merging. Such pipelines streamline the quality assurance process and increase productivity. However, current techniques in static analysis are limited: they are prone to false positives (wasting developer effort), and false negatives (making the analysis less reliable)~\cite{1628970, anderson2008use}.

In recent years, there has been an increase in the application of statistical models, namely neural networks, to a variety of code intelligence tasks, including vulnerability detection~\cite{harer2018automated, 9108283}. This research has mainly focused on the application of pre-trained models that capture knowledge particular to a specific programming language, and has been inspired by transformer-based models~\cite{NIPS2017_3f5ee243} such as BERT~\cite{devlin-etal-2019-bert} and GPT~\cite{radford2018improving}, both developed in the context of natural language processing (NLP). For example, models such as CodeBERT~\cite{feng2020codebert}, C-BERT~\cite{buratti2020exploring} and PLBART~\cite{ahmad2021unified} produce distributed representations from source code that have been applied to many code tasks, such as code search, code translation and vulnerability detection. The recent CodeXGLUE benchmark~\cite{lu2021codexglue} aims to facilitate the comparison and evaluation of these recent models in a large variety of tasks. This benchmark is publicly available and open for for further contributions \footnote{\url{https://github.com/microsoft/CodeXGLUE}}.

Although these models are demonstrably effective, they also have their limitations: mainly, large models with hundreds of millions of parameters need a relatively large amount of computational resources, including both memory and CPU time \cite{sohoni2019low}. The requirement for large amounts of computational resources is a significant limitation for researchers and developers, and it can make the usage of large pre-trained models impractical. There is a noticeable trade-off between model representativeness and convenience of use. As a result, often smaller models are preferred for detection performance. 
% As such, it is helpful to use smaller models to achieve a satisfactory result even if it is not up to state-of-the-art standards.

\begin{figure*}[t!]
\centering
\begin{minipage}[b]{0.5\linewidth}
\begin{minted}[fontsize=\scriptsize]{c}
void scsi_req_abort(SCSIRequest *req, int status) {
    if (!req->enqueued) {
        return;
    }
    scsi_req_ref(req);
    scsi_req_dequeue(req);
    req->io_canceled = true;
    if (req->ops->cancel_io) {
        req->ops->cancel_io(req);
    }
    scsi_req_complete(req, status);
    scsi_req_unref(req);
}
\end{minted}
\caption{Example of non-vulnerable function.}
\label{fig:function1}
\vspace{1cm}
\end{minipage}
\quad
\begin{minipage}[b]{0.5\linewidth}
\begin{minted}[fontsize=\scriptsize]{c}
uint32_t HELPER(shr_cc)(CPUM68KState *env, uint32_t val, uint32_t shift) {
    uint64_t temp;
    uint32_t result;
    shift &= 63;
    temp = (uint64_t)val << 32 >> shift;
    result = temp >> 32;
    env->cc_c = (temp >> 31) & 1;
    env->cc_n = result;
    env->cc_z = result;
    env->cc_v = 0;
    env->cc_x = shift ? env->cc_c : env->cc_x;
    return result;
}
\end{minted}
\caption{Example of vulnerable function.}
\label{fig:function2}
\end{minipage}
\label{fig:functions}
\end{figure*}

In this work, we investigated the applicability of a code representation model, Code2vec~\cite{10.1145/3290353}, to the vulnerability detection task. Code2vec is built on a simple attention-based feed-forward neural network that learns and combines semantic knowledge extracted from AST path-contexts. To the best of our knowledge, Code2vec has not been evaluated for vulnerability detection. We evaluated Code2vec on a labeled corpus of C functions. We found that Code2vec outperformed traditional NLP-based approaches for vulnerability detection at an accuracy of $61.43\%$, comparable to simple transformer models such as pre-trained RoBERTa~\cite{liu2019roberta}. We show that these results are achievable at a fraction of the computational resources and training time necessary for transformer-based models; Code2vec ran a full training session in approximately $5$ minutes on a consumer-grade GPU, handling $1024$ samples of data at each training step without exhausting its memory. We submitted our results to the CodeXGLUE leaderboard for the defect detection task for comparison with the state-of-the-art.

Our contributions are as follows:

\begin{itemize}
    \item An evaluation of Code2vec on the task of vulnerability detection using a dataset of labeled C functions both regarding accuracy and computational requirements (training time and memory).
    \item A replication package with the scripts and data to train and test the model, for reproducibility. Available online: \url{https://github.com/dcoimbra/dx2021}.
\end{itemize}

The paper is organized as follows: in section~\ref{sec:approach}, we present our approach for applying Code2vec to vulnerability detection. In section~\ref{sec:implementation}, we describe the implementation details related to our extraction of path-contexts and Code2vec configuration. In section~\ref{sec:evaluation}, we describe the evaluation metrics employed and present our results alongside previous studies, and discuss them in section~\ref{sec:discussion}. In section~\ref{sec:rw}, we give a brief summary of the related work in deep learning for vulnerability detection. Finally, section~\ref{sec:conclusion} presents our conclusions and lays discusses future work.

\section{Approach}
\label{sec:approach}

In this section, we describe our approach for using path-context representations for the detection of security vulnerabilities using Code2vec. We describe our procedure for the extraction of path-contexts from a corpus of labeled functions and provide a summary of Code2vec and how we adapted it for the vulnerability detection task. 

\subsection{Dataset}

\begin{table}[t]
    \centering
    \begin{tabular}{|c|c|c|}
\hline
      & Vulnerable  & Non-Vulnerable \\ \hline\hline
Train & $10018$ & $11836$    \\ \hline
Test  & $1255$  & $1477$     \\ \hline
Validation   & $1187$  & $1545$     \\ \hline
\end{tabular}
    \caption{Distribution of vulnerable and non-vulnerable functions in Devign.}
    \label{tab:dataset}
\end{table}

We used the public dataset Devign\footnote{\url{https://sites.google.com/view/devign}}~\cite{NEURIPS2019_49265d24}, provided as part of the CodeXGLUE benchmark. Devign includes $27318$ manually-labeled functions collected from QEMU and FFmpeg, two large C open-source projects. These functions were extracted by collecting security-related commits and selecting vulnerable and non-vulnerable versions of functions from the labeled commits. Each function was manually labeled by a group of three professional security experts. Functions are labeled as $1$ (vulnerable) and $0$ (non-vulnerable), with no distinction made with regard to the type of vulnerability. Examples of a non-vulnerable and vulnerable functions extracted from Devign are displayed in Figures~\ref{fig:function1} and~\ref{fig:function2} respectively. We used the dataset splits as prepared by CodeXGLUE\footnote{\url{https://github.com/microsoft/CodeXGLUE/tree/main/Code-Code/Defect-detection}}: $80\%$/$10\%$/$10\%$ for training, validation, and testing, respectively. The distribution of vulnerable and non-vulnerable functions for the dataset is presented in Table~\ref{tab:dataset}.

\begin{figure*}[t!]
    \centering
    \includegraphics[width=0.6\linewidth]{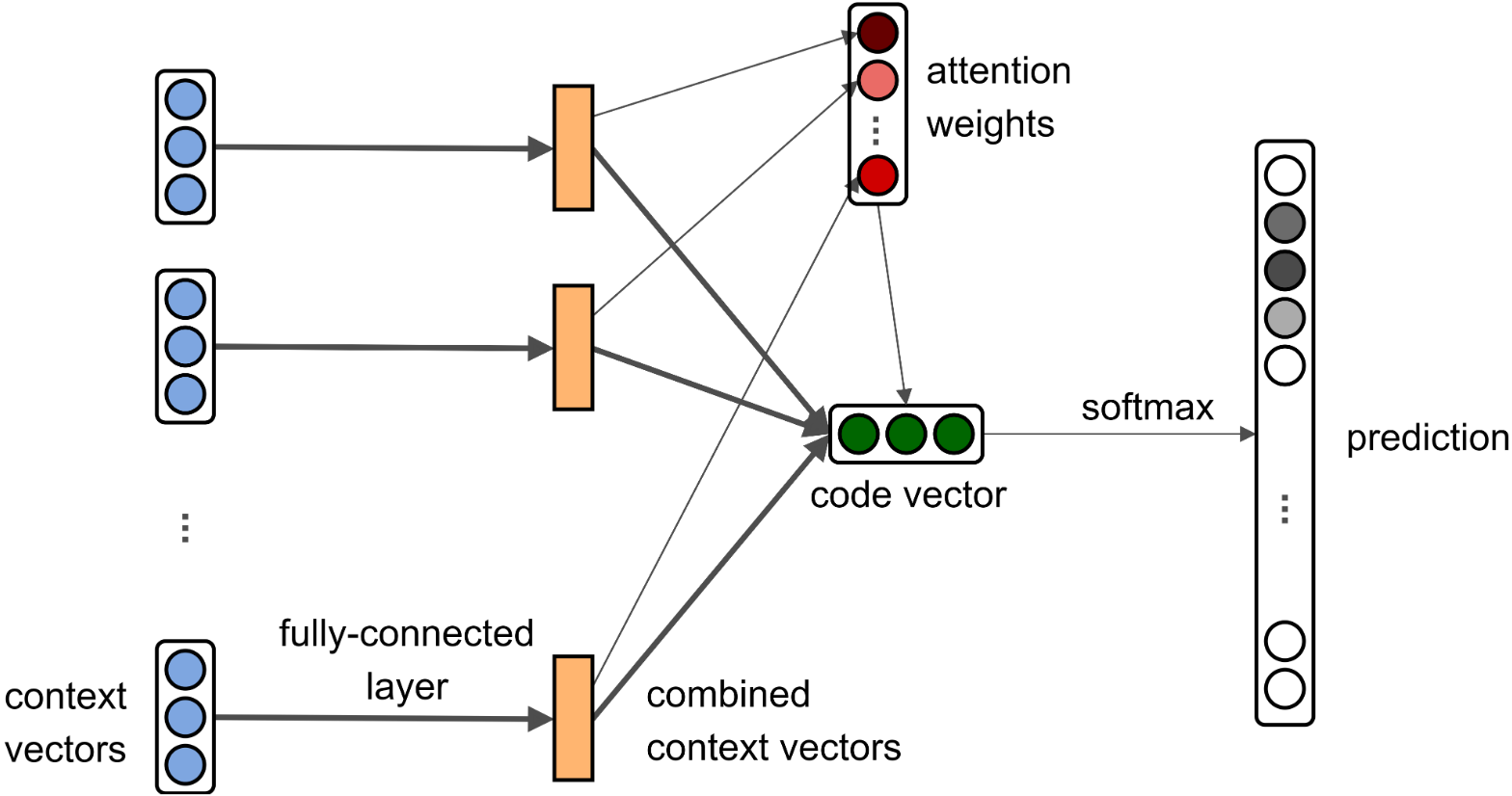}
    \caption{Code2vec architecture. Adapted from original image by~\cite{10.1145/3290353}}
    \label{fig:Code2vec}
\end{figure*}

\subsection{Representing code snippets as bags of path-contexts}

The path-attention model on which Code2vec is built takes a source code function as represented by a set of \emph{path-contexts}, extracted from its abstract syntax tree (AST). We describe the definitions of AST, AST path and path-context as presented in the Code2vec paper:

\begin{definition}[Abstract Syntax Tree]
An abstract syntax tree (AST) for a code snippet $\mathcal{C}$ is a tuple $<N, T, X, s, \delta, \phi>$ where $N$ is a set of non-terminal nodes, $T$ is a set of terminal nodes, $X$ is a set of values, $s \in N$ is the root node, $\delta : N \xrightarrow{} (N \cup T)^*$ is a function that maps a non-terminal node to a list of its children, and $\phi : T \xrightarrow{} X$ is a function that maps a terminal node to its associated value. Every node except the root appears exactly once in all the lists of children; that is, each node has exactly one parent. 
\end{definition}

\begin{definition}[AST path]
An AST path of length $k$ is a sequence of the form: $n_1d_1 \dots n_kd_kn_{k+1},$ where $n_1,n_{k+1} \in T$ are terminal nodes,  $\forall{i} \in [2..k]: n_i \in N$ are non-terminal nodes and $\forall{i} \in [1..k]: d_i \in \{\uparrow,\downarrow\}$ are movement directions (up or down in the tree). If $d_i = \uparrow$, then $n_i \in \delta(n_{i+1})$; if $d_i = \downarrow$, then $n_{i+1} \in \delta(n_i)$.
\end{definition}

\begin{definition}[Path-context]
Given an AST path $p$, a path-context is a triplet $<x_s, p, x_t>$ where $p$ is a syntactic path in the AST and $x_s$ and $x_t$ correspond to the values associated with the start and end terminals of $p$, respectively. A possible path-context for the statement \texttt{x = 7} would be:
\begin{equation*}
\scriptstyle
    <\texttt{x}, (NameExpr \uparrow AssignExpr \downarrow IntegerLiteralExpr), \texttt{7}>
\end{equation*}
\end{definition}

It is possible to limit the paths by a maximum length and a maximum width (distance between two child nodes of the same intermediate node). A code snippet $\mathcal{C}$ is represented as a \emph{bag of path-contexts} consisting of path-contexts extracted from the AST for $\mathcal{C}$. We kept the Code2vec defaults of maximum length $8$ and maximum width $3$ and, for each function in the corpus, extracted a bag of at most $200$ path-contexts.

\subsection{The Code2vec model}

Code2vec learns embedding matrices for paths, values and labels ($\boldsymbol{path\_vocab}$, $\boldsymbol{value\_vocab}$, $\boldsymbol{tags\_vocab}$, respectively), a fully-connected layer, and an attention vector $\boldsymbol{a}$. An illustration of the Code2vec architecture is shown in Figure~\ref{fig:Code2vec}. An embedding for a single path-context $b_i = <x_s, p_j, x_t>$ is a \emph{context vector} $\boldsymbol{c_i}$ which corresponds to the concatenation of the embeddings of the start and end tokens and of their connecting paths:

\begin{equation}
    \begin{aligned}[b]
    \boldsymbol{c_i} = \mathrm{embedding}(<x_s,p_j,x_t>) = \\     
    \scriptstyle [\boldsymbol{value\_vocab}_s ; \boldsymbol{path\_vocab}_j ; \boldsymbol{value\_vocab}_t] 
    \in \mathbb{R}^{3d}
    \end{aligned}
\end{equation}

In the previous equation, the operator $;$ is the concatenation operator and $d$ is an empirically-determined hyperparameter defining the length of the internal representation.

A fully connected layer of Code2vec learns to combine each component of the embedding of a path-context, through a simple linear combination with a learned weights matrix $\boldsymbol{W}$ passed through a hyperbolic tangent function:

\begin{equation}
    \boldsymbol{\Tilde{c_i}} = \tanh(\boldsymbol{W} \cdot \boldsymbol{c_i}) \in \mathbb{R}^d
\end{equation}

In the previous equation, $\boldsymbol{W} \in \mathbb{R}^{3d \times d}$. Finally, Code2vec's attention mechanism aggregates all combined context vectors $\{ \boldsymbol{\Tilde{c}_1},\dots,\boldsymbol{\Tilde{c}_n}\}$ into a single representation. The attention weight $\alpha_i$ of each $\boldsymbol{\Tilde{c}_i}$ is obtained through the normalized inner product between $\boldsymbol{\Tilde{c}_i}$ and the global attention vector $\boldsymbol{a}$.

\begin{equation}
    \alpha_i = \frac{\mathrm{exp(\boldsymbol{\Tilde{c}_i^\intercal} \cdot \boldsymbol{a})}}{\sum^n_{j=1} \mathrm{exp(\boldsymbol{\Tilde{c}_j^\intercal} \cdot \boldsymbol{a})}}
\end{equation}

The final code vector $\boldsymbol{v}$ is a weighted average of the combined context vectors factored by their attention weights:

\begin{equation}
    \boldsymbol{v} = \sum^n_{i = 1} \alpha_i \cdot \boldsymbol{\Tilde{c}_i}
\end{equation}

For prediction, the probability that a specific label $y_i$ is assigned to a code snippet $\mathcal{C}$ is the normalized inner product between the embedding for $y_i$ and the code vector $\boldsymbol{v}$:

\begin{equation}
    \forall{y_i} \in Y : q(y_i) = \frac{\mathrm{exp}(\boldsymbol{v^\intercal} \cdot \boldsymbol{tags\_vocab}_i)}{\sum_{y_j \in Y} \mathrm{exp}(\boldsymbol{v^\intercal} \cdot \boldsymbol{tags\_vocab}_j)}
\end{equation}

In the previous equation, $Y$ is the set of label values found in the training corpus. Training is performed by minimizing cross-entropy loss using the Adam optimization algorithm. For inference, we take the target label to which Code2vec assigned the highest probability.

\section{Implementation}
\label{sec:implementation}

\begin{figure*}[t!]
    \centering
    \includegraphics[width=0.8\linewidth]{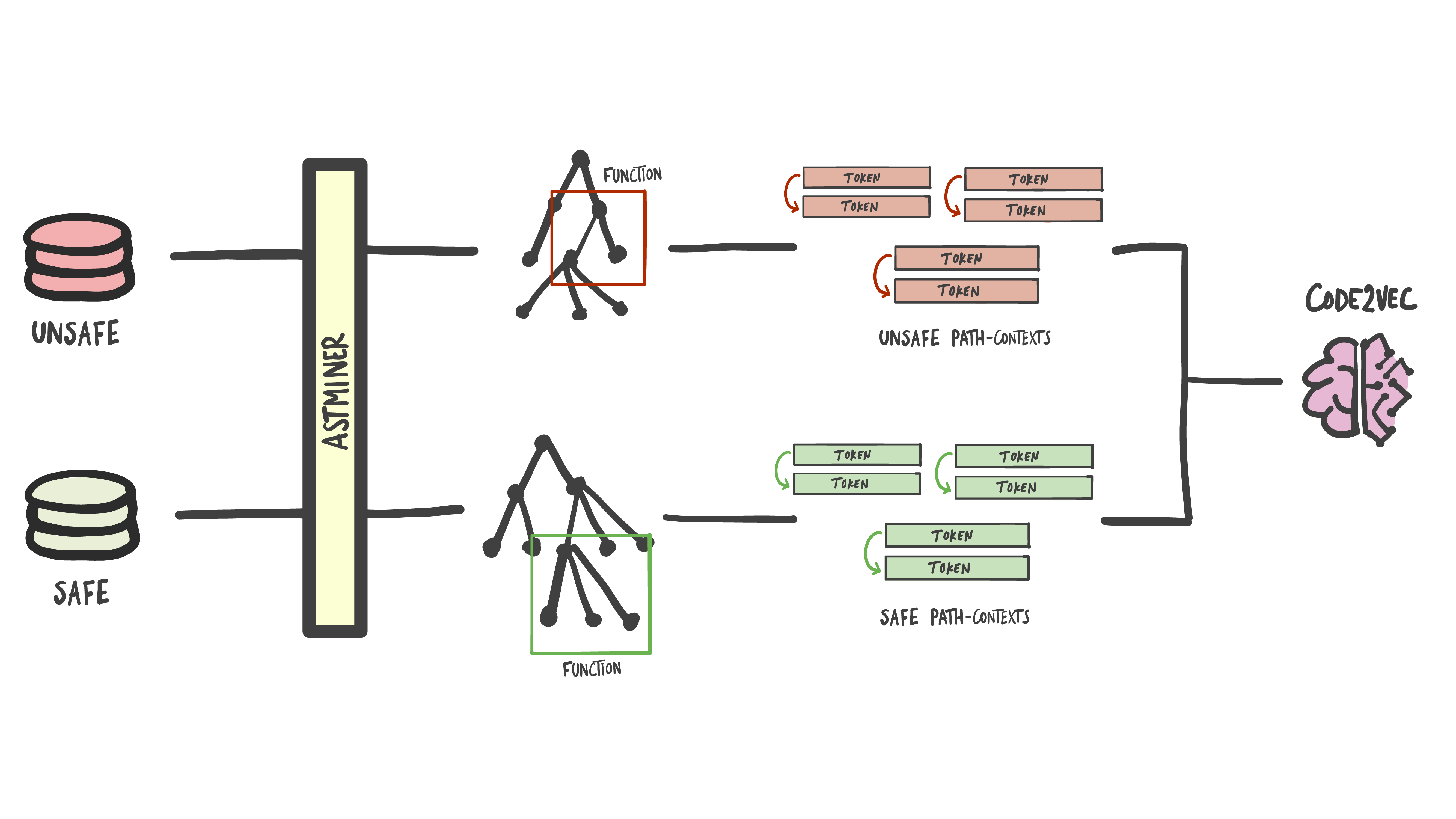}
    \vspace{-1cm}
    \caption{Experiments pipeline. The vulnerable and non-vulnerable raw source code functions are independently passed to astminer. For each function, astminer extracts the respective AST, from which it computes an appropriate bag of path-contexts. After labeling each bag of path-contexts corresponding to each function, the result is passed to Code2vec for training. The trained model is then used for inference.}
    \label{fig:pipeline}
\end{figure*}

In this section, we describe the implementation details for our approach. We describe the open-source library astminer, which we leveraged for extracting bags of path-contexts, as well as the parameters used for extraction and training. Our final pipeline is illustrated in Figure~\ref{fig:pipeline}.

\subsection{Extracting bags of path-contexts}

To extract a set of path-contexts for each code snippet in Devign, we use the open-source library astminer\footnote{\url{https://github.com/JetBrains-Research/astminer}}~\cite{kovalenko2019pathminer}. We wrote a custom script that visits each function in the corpus, and computes its path-contexts. As Code2vec is designed to predict the names of functions rather than numeric labels, our script converts each label of $0$ or $1$ in the original dataset to the string tokens ```safe''' or ```vuln''' respectively. Following Code2vec defaults, we limit the length and width of each path-context to $8$ and $3$ respectively, and extract at most $200$ path-contexts per function.

By default, astminer replaces each value and path in a path-context with a corresponding ID number in order to reduce memory consumption and training time when passing the samples to Code2vec. This is implemented by maintaining tables of $<\mathrm{id}, \mathrm{value}>$ pairs for tokens, node types, and paths throughout the entire run. However, this was considerably memory-intensive on the machine we used. As such, we bypassed this feature and directly extracted path-contexts in their original format. We computed the MD5 hash of the string representation of the path component of each path-context instead of extracting the entire path as-is. This process is identical to the one used in the Code2vec paper for function name prediction~\cite{10.1145/3290353}.

\begin{table}[t]
    \centering
    \begin{tabular}{|c|c|c|}
\hline
      & Vulnerable  & Non-Vulnerable \\ \hline\hline
Train & 9987 & 11809    \\ \hline
Test  & 1253  & 1472     \\ \hline
Validation   & 1185  & 1541     \\ \hline
\end{tabular}
    \caption{Distribution of vulnerable and non-vulnerable functions in Devign after applying astminer.}
    \label{tab:datasetminer}
\end{table}

The astminer library was unable to extract path-contexts from $71$ functions of the Devign dataset. Therefore, these samples were not included in our study. The final distribution of functions after applying astminer is described in Table~\ref{tab:datasetminer}: $9987$ vulnerable functions and $11809$ non-vulnerable functions for the training dataset; $1253$ vulnerable functions and $1473$ non-vulnerable functions for the testing dataset;
and, finally, $1185$ vulnerable functions and $1541$ non-vulnerable functions for the validation dataset.

\subsection{Code2vec}

We used the official implementation of Code2vec\footnote{\url{https://github.com/tech-srl/Code2vec}}. We trained for $20$ epochs and performed inference on the epoch with the highest F1-score on the validation dataset. The performance measures of the model were 
adapted to the vulnerability detection task: we considered a prediction of ```safe''' to be a negative prediction while a prediction of ```vuln''' to be a positive prediction. The training hyperparameters followed Code2vec defaults: batch size of $1024$, embedding size of $128$, and dropout rate of $0.25$.

\section{Evaluation}
\begin{table}[t]
    \centering
\begin{tabular}{|c|c|c|c|c|}\hline
Model             & Accuracy       & Precision      & Recall         & F1             \\ \hline\hline
CoTexT            & $66.62$          & -              & -              & -              \\ \hline
C-BERT            & $65.45$          & -              & -              & -              \\ \hline
PLBART            & $63.18$          & -              & -              & -              \\ \hline
CodeBERT          & $62.08$          & -              & -              & -              \\ \hline
\textbf{Code2vec} & \textbf{61.43} & \textbf{57.50} & \textbf{61.77} & \textbf{59.56} \\ \hline
RoBERTa           & $61.05$          & -              & -              & -              \\ \hline
TextCNN           & $60.69$          & -              & -              & -              \\ \hline
BiLSTM            & $59.37$          & -              & -              & -              \\ \hline
\end{tabular}
    \caption{Results for the Devign dataset alongside the CodeXGLUE leaderboard. Our contributions are in bold.}
    \label{tab:results}
\end{table}

\label{sec:evaluation}
\begin{table}[t]
\centering
\begin{tabular}{|c|c|c|c|c|}
\hline
Model & Train Time & \#Epochs & Hardware                         \\ \hline\hline
\textbf{Code2vec} & \textbf{5 minutes} & \textbf{20} & \textbf{1050Ti x1} \\ \hline
\textbf{CodeBERT} & \textbf{7 hours} & \textbf{5} & \textbf{1050Ti x1} \\\hline
CodeBERT & 1 hour & 5 & Tesla P100 x2 \\ \hline
\end{tabular}
\caption{Training time on the Devign dataset alongside the CodeXGLUE measurement, for a complete training session. Our contributions are in bold.}
\label{tab:time}
\end{table}

\begin{table}[t]
\centering
\begin{tabular}{|c|c|c|c|}
\hline
Model & \#Params & Embedding size & Memory (MB)                         \\ \hline\hline
Code2vec & 31M & $128$ & $600$ \\ \hline
CodeBERT & 125M & $400$ & $2484$ \\\hline
\end{tabular}
\caption{Parameter count and memory consumption for each model. Memory consumption is defined as the amount occupied in RAM by the model and a single sample of data during a training step, with all gradients loaded.}
\label{tab:memory}
\end{table}

The CodeXGLUE benchmark for defect detection reports only accuracy as an evaluation metric. As Devign is a balanced dataset, accuracy is an appropriate metric for assessing performance. Nevertheless, we also computed precision, recall and F1-score in addition to accuracy. These metrics help us assess the model's ability to distinguish between vulnerable and non-vulnerable samples.  Our results are shown in Table~\ref{tab:results} alongside the current entries in the CodeXGLUE leaderboard. We note that CoTexT had not been released at the time of writing. In addition to these scores, we compared the training time and memory consumption for Code2vec and CodeBERT on this task on our hardware. We computed training time for a complete training session for each model, which corresponds to $20$ epochs on Code2vec and $5$ epochs on CodeBERT. We chose $5$ epochs for CodeBERT as this is the value used in CodeXGLUE to obtain the original results. Training times for each model are presented in Table~\ref{tab:time} while memory consumption is shown in Table~\ref{tab:memory}.

\section{Results and Discussion}
\label{sec:discussion}

Based on accuracy alone, Code2vec outperformed the traditional NLP-based methods BiLSTM~\cite{GRAVES2005602} ($61.43 > 59.37$) and TextCNN~\cite{kim-2014-convolutional} ($61.43 > 60.69$). The obtained accuracy score for Code2vec was slightly higher than pre-trained RoBERTa ($61.43 > 61.05$): the two  methods have similar performance. This is to be expected as these models were not designed for code intelligence tasks, nor pre-trained on source code data. In the same vein, Code2vec was outperformed by PLBART ($61.43 < 63.18$), C-BERT ($61.43 < 65.45$) and CodeBERT ($61.43 < 62.08$), which uses the transformer architecture to learn source-code features through pre-training on large amounts of source code data. As an advantage, transformer-based models do not require an intermediate representation and can be fine-tuned on source code directly. 

Regarding training time, as shown in Table~\ref{tab:time}, Code2vec completed a $20$-epoch training session in approximately $5$ minutes on our hardware. Conversely, on the same hardware, CodeBERT completes a $5$-epoch training session in approximately $7$ hours. This is to be expected, as transformer-based models need a larger amount of parameters, in turn requiring much more computational resources to process the large amounts of data. Additionally, transformer-based models create internal representations that occupy large amounts of memory, which typically are not available on consumer-grade hardware: therefore, training has to be carried out in very small batches of data, increasing training time. A large advantage of Code2vec compared with transformer-based models is its relatively low memory footprint: as shown in Table~\ref{tab:memory}, a single training step with CodeBERT requires approximately $2.5$GB of memory to complete one training step, an amount mostly represented by saved gradients during back-propagation. Conversely, a single training step with Code2vec was performed with just $600$MB of GPU memory. This is due to Code2vec's simpler architecture, allowing for a lower amount of gradients to be saved during training, as well as a lower number of parameters and smaller embedding sizes. Additionally, Code2vec's lower memory footprint allows us to load larger batches of data into memory at each training step, increasing training performance. 

\section{Related Work}
\label{sec:rw}

Recent research on machine learning 
for security vulnerability detection has used both token-based and graph-based approaches ~\cite{Li_2018, 9321538, russell2018automated}.

Token-based models consider the code as a sequence of tokens. Several models
have been proposed using different neural network architectures such as Bidirectional
Long-Short-Term Memory (BSLTM)~\cite{Li_2018}, Convolutional Neural Networks
(CNN)~\cite{russell2018automated}, and Recurrent Neural Networks (RNN).
However, simple token-based models struggle to reason about the long sequences produced
from transforming source code into token sequences.
To help address this problem, newer approaches using code slices instead of the entire code sequences were proposed, see for example  VulDeePecker~\cite{Li_2018} 
and SySeVR~\cite{9321538}. The hypothesis behind slicing is that 
different parts of the code are not equally important for 
the model to learn vulnerability patterns. Therefore, 
these newer approaches consider only slices extracted 
from \emph{interesting points} (e.g., API calls)---points
considered important for vulnerability prediction. The 
rest of the code elements are ignored.
Token-based approaches usually fail to maintain the dependencies 
between the tokens that are the root of the problem. Thus, learning 
those dependencies (or semantic relationships) is at best difficult and at worst impossible.

Graph-based models incorporate syntactic and semantic dependencies
between different code elements. Source code can be transformed 
to syntactic graphs (Abstract Syntax Trees) and semantic graphs (Control
Flow Graphs, Data Flow Graphs, Program Dependency Graphs, and so on).
Devign~\cite{NEURIPS2019_49265d24} uses Code Property Graphs (CPGs) to build 
a vulnerability detection model as proposed by Yamaguchi et al.~\cite{6956589}.
Chakraborty et al. ~\cite{chakraborty2020deep} also generate code property graphs from source code
to consider the syntax and the semantics of the code. CPGs consider succinct information
from the control-flow and data-flow graphs in addition to the AST and the program
dependency graphs. Each of these elements offer additional context about the 
semantic structure of the code that may be relevant for vulnerability detection.

Both token and graph-based models suffer from vocabulary explosion---the number
of possible identifiers (e.g., variable and function names) in the code can be infinite. Some approaches replace tokens
with abstract names~\cite{Li_2018, 9321538}. Other techniques use
word embedding tools (e.g., word2vec) to create vector representations
of every token. For instance, VulDeePecker~\cite{Li_2018} and SySeVR~\cite{9321538} use word2vec to transform 
symbolic tokens into vectors. In contrast, Devign~\cite{NEURIPS2019_49265d24} uses word2vec to transform 
 pure code tokens to real vectors.
Alon et al.~\cite{10.1145/3290353} proposed continuous
distributed vector (or code embeddings) to represent code. Code2vec aggregates an arbitrary sized snippet of code
into a fixed-size vector in a way that captures its semantics. Code functions are 
transformed in groups of path-contexts, where each path-context represents a semantic 
relationship between two code elements in a function. 

CodeBERT~\cite{feng2020codebert} is a transformer-based model that represents given snippets of source code in a distributed representation vector~\cite{NIPS2017_3f5ee243}. The non-sequential nature of the architecture of the transformer encoder, being based on a simple attention mechanism, is designed to address the problem of reasoning about long sequences: each token is processed in parallel throughout the model. CodeBERT was pre-trained on pairs of programming language and natural language data. 
% Pre-training is based on two tasks: masked language modeling (where programming language and natural language tokens are randomly masked and the model is asked to predict those tokens) and replaced token detection (where tokens are randomly replaced with plausible substitutions, and the model is asked to predict whether the token is the original one or a substitution). 
A pre-trained model produces distributed representations that can be used in a variety of downstream tasks, on which the model itself can be further fine-tuned.  

This study evaluated how Code2vec---a model that considers syntactic and semantic 
relationships in the code---fairs compared to  other non-token based and token-based models (specifically, CodeBERT) for vulnerability detection in C/C++.

\section{Conclusions}
\label{sec:conclusion}
We applied Code2vec, a model for distributed code representations using AST path-contexts, to the task of binary vulnerability detection. We evaluated Code2vec on the Devign dataset as part of the CodeXGLUE benchmark. Our experiments achieved an accuracy score that outperformed naive NLP-based approaches and was equivalent to a simple transformer-based model that had not been pre-trained on source code data. However, as expected, it was outperformed by more expressive models that directly leverage features unique to the source code. Additionally, we showed that smaller models such as Code2vec have modest computational resource requirements compared to other alternatives: when computational resources are scarce, the reduced training time requirement and memory consumption of Code2vec on a labeled dataset of source code functions may outweigh the loss in accuracy that results from its lower expressiveness. 

The performance of Code2vec for vulnerability detection could potentially be improved through hyperparameter tuning, for example, by optimally choosing the maximum length and width of path-contexts as well as the learning rate and batch size. We plan to investigate these improvements in future work. Although Code2vec is limited %in representativeness
%corina: not sure representativeness is a word
compared to state-of-the-art, it remains an attractive choice for developers for code intelligence tasks such as vulnerability detection, as it demonstrates comparable performance and can be more easily integrated in CI pipelines than static analyzers or larger models.

\section*{Acknowledgments}
This work was supported in part by Fundação para a Ciencia e a Tecnologia (FCT) 
under Grants CMU/TIC/0064/2019 (a project funded by the Carnegie Mellon Portugal Program) 
and UIDB/50021/2020.

%% The file dx.bst is a bibliography style file for BibTeX 0.99c
% \balance
\bibliographystyle{unsrt}
\bibliography{dx}

\end{document}